# Theoretical Modeling Expressions for Networked Enzymatic Signal Processing Steps as Logic Gates Optimized by Filtering

Vladimir Privman

*Department of Physics, Clarkson University, Potsdam, NY 13699*

**Abstract**

We describe modeling approaches to a "network" of connected enzyme-catalyzed reactions, with added (bio)chemical processes that introduce biochemical filtering steps into the functioning of such a biocatalytic cascade. Theoretical expressions are derived that allow simple, few-parameter modeling of processes concatenated in such cascades, both with and without filtering. The modeling approach captures and explains features identified in earlier studies of enzymatic processes considered as potential "network components" for multi-step information/signal processing systems.

# INTRODUCTION

Concatenated biomolecular reactions allow development of tailored-response[1-6] and complex signal processing[7,8] systems for multi-input biosensing[9-17] and for information processing[18-29] without electronics, the latter termed "biomolecular computing" or "biocomputing." Avoidance of standard electronic components can in some situations offer interesting new functionalities and applications.[30-35] Furthermore, the output, as well as the inputs and other process steps can be signals that involve interfacing with electronics[36-44] or other signal-responsive materials.[45-51] Recent studies resulted in improvement of linear response of biosensors.[3] Another development has involved obtaining sigmoid response of certain biocomputing "gates" by chemical modifications of enzymatic processes.[4-6,52-56] The latter approach improves detection of biomarker combinations for medical diagnostics.[9-17] Small model networks of biochemical steps for biocomputing have also been considered.[18-29] Approaches to optimizing individual biochemical processes' as "network components" (gates, etc.) and whole networks' functioning to avoid noise amplification have been reported.[22,31-33,57-59]

Biocomputing[23,24,59-61] is a subfield of "logic" chemical systems[62-66] and then in turn of unconventional computing.[67,68] Biocomputing utilizes biomolecules: proteins/enzymes,[23,24,69,70] DNA,[27,28,30,71] RNA[72,73] and even whole cells,[74,75] which offers specificity and selectivity, enabling networking relatively complex systems without reaction cross-talk of the processes involved. Enzyme-based systems are of special interest in biosensing applications[9-17] and can be integrated with electronic devices[36-44] and signal-responsive materials.[45-51]

The advance of experimental realizations has also required new theoretical modeling ideas[52-59] to allow few-parameter description of various biochemical and chemical reaction processes included in information/signal processing cascades. There have been[22,59,76] studies that tested modeling approaches for networked biochemical steps. However, the latest ideas of biochemical filtering,[4-6,52-56,77-79] usually accomplished by adding chemical or enzymatic reactions to enzyme-catalyzed processes that are the "gates" in digital signal processing, have only been confronted with modeling relatively recently.[59] Here we review the theoretical results that were developed[59] to model such few-step networks of connected biochemical signal processing steps, with and without added filtering reactions.

**GATES AND FILTERING STEPS IN CASCADES OF BIOMOLECULAR PROCESSES**

Let us consider a cascade of biomolecular processes as a model to explore ideas of parameterizing and optimizing the functioning as such systems as small networks for signal processing and biocomputing. This process cascade was designed in Ref. 59, and its constituent processes are shown in Figure 1. In fact, this cascade consists of steps similar to those which have also been incorporated in enzymatic biosensors involving detection of maltose or starch.[80-84]

The first step functions as an AND logic gate. Its two inputs are maltose (selected as logic Input 1) and phosphate (Input 2). The interpretation of this and other process steps as binary "logic gates" and also non-binary (analog) "filtering" functions is also shown in Figure 1, and we will further discuss this below. The output, glucose, of this gate is an input for the reaction steps biocatalyzed by enzyme GOx (see Figure 1), resulting in the production of $H_2O_2$. This will be considered an identity (I) binary "gate." $H_2O_2$ in turn is an input for enzyme HRP processes, which also use TMB. The latter is selected as logic Input 3. The output of this AND function is $TMB_{ox}$. The final output signal can be measured optically as described in Ref. 59. This optical measurement of the $TMB_{ox}$ concentration can be viewed as another I-gate step in the network.

Two different "filtering" processes, marked by F, can be also carried out, singly or together, one biocatalyzed by enzyme HK, the other involving the "recycling" of the output chemical by its reacting with NADH. These non-binary processing steps are shown in our "network" interpretation in Figure 2. We will revisit each of the binary and non-binary network steps separately later. One of the interesting features of this network is that it can be made more complicated[59] by utilizing certain known[85-88] enzymatic and chemical processes and properties.

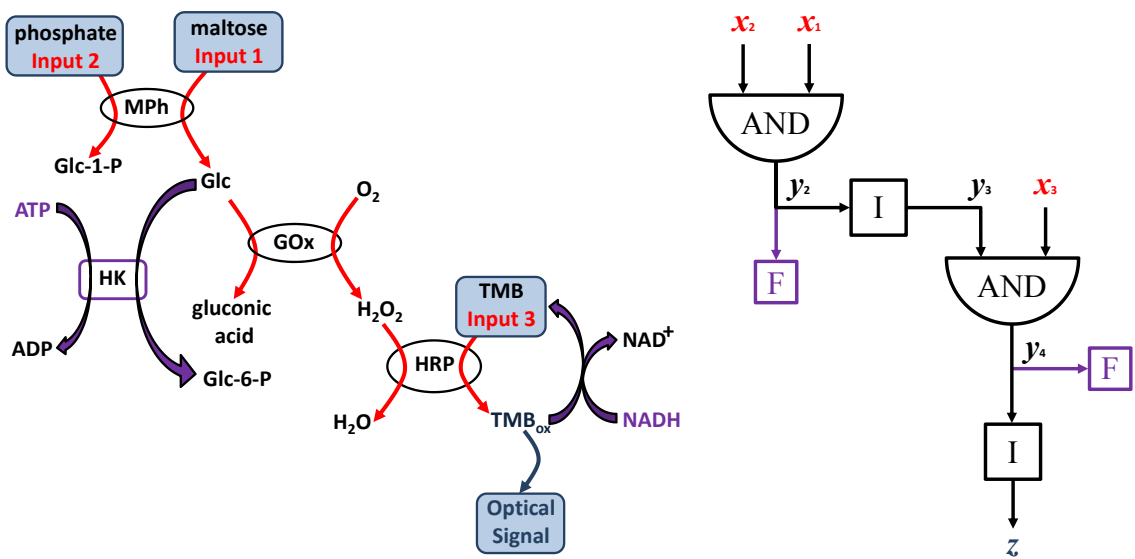

**Figure 1.** Left: The biocatalytic cascade with three variable inputs: maltose, phosphate, 3,3',5,5'-tetramethylbenzidine (TMB), and two optional added "filtering" processes, one biocatalyzed by hexokinase (HK), and another involving the "recycling" of the output chemical, the oxidized TMB (TMB$_{ox}$) by NADH, the latter the reduced cofactor β-nicotinamide adenine dinucleotide (NAD$^{+}$). Abbreviations for various chemicals are as follows: maltose phosphorylase (MPh), glucose oxidase (GOx), horseradish peroxidase (HRP), adenosine 5'-triphosphate (ATP) and adenosine diphosphate (ADP), glucose (Glc), β-D-glucose-1-phosphate (Glc-1-P) and α-D-glucose-6-phosphate (Glc-6-P). Right: The (bio)chemical processes shown can be viewed as a "network" of binary AND and identity (I) gates, with added non-binary "filtering" (F) functions, as explained in the text.

The response of the biocatalytic cascade to the variation of the initial "signal" concentrations of enzymes' substrates selected as inputs (Figure 1), maltose, phosphate and TMB, is measured with inputs starting at concentration 0, as the reference logic-**0** values, and increasing up to conveniently selected reference logic-**1** values, the latter typically, in the order of magnitude of 1 to 10 mM range depending on the application. Concentrations of the "gate machinery" chemicals, MPh, GOx, HRP also control the system's response, but only to a limited extent.[22] Instead, as reviewed below, the gate performance is optimized by the added filter process. The appropriate chemicals, HK and ATP, activate the HK-filter, and NADH activates NADH-filter (Figure 1).

Enzymatic processes considered here, such as those in Figure 1, typically have complicated mechanisms of action, some not fully known. For example, for enzyme MPh the specifics of the mechanism of its biocatalytic steps are not well studied, and the order of intake of the two substrates is not unique.[89,90] In the next section, we describe the motivation for a modeling approach suitable for evaluation of such systems as logic-gate networks. Note that the designation of the "logic inputs," such as maltose and phosphate, out Inputs 1 and 2, for information processing is made based on the desired application and does not imply that this is the actual kinetic order of their intake. At time $t = 0$, the inputs are varied from some application-determined logic-**0** values, here taken as the initial concentrations 0 for simplicity, to logic-**1** values. For analysis of the system's functioning as a logic network, we then define scaled variables in the range from **0** to **1**, here, for example,

$$x_1 = [\text{maltose}](t = 0)/[\text{maltose}]_{\max}, \tag{1}$$

where $t$ denotes the time, and $[\text{maltose}]_{\max}$ is the maximum (logic-**1**) initial ($t = 0$) concentration for Input 1. Variables $x_2$ and $x_3$ are defined similarly. For the output signal, we define

$$z = \text{Abs}_{\mathbf{000}}(t = t_g)/\text{Abs}_{\mathbf{111}}(t = t_g), \tag{2}$$

where the absorbance, Abs, of the generated chemical output signal $\text{TMB}_{\text{ox}}$ is measured at the gate time, $t_g > 0$. The definition of the logic variables $y_{2,3,4}$ for the intermediate products, see Figure 1, are also similar, but for $y_{2,3}$, in particular, they require additional discussion because of time dependence. We will address this later.

The reference logic values of the output should be set by the system functioning: logic-**0** at zero inputs, **000**, and also for inputs **001**, **101**, etc., totaling seven combinations with at least one zero, and logic-**1** at inputs **111**. This will be accurate only if the system actually functions as expected, i.e., as the two-AND gate network shown in Figure 1. Deviations from precise binary gate performance are exactly the reason for considering the gate-functions' "quality," effects of noise buildup, and the gates' network response to variation of the inputs in and somewhat beyond the [0,1] "logic" ranges, rather than just focusing at the binary logic points **0** and **1**.

In modeling networks of the type considered here, we seek to evaluate their utility as information processing systems. Therefore, an approximate few-parameter description suffices in most cases to parameterize the "response shape," i.e., the function $z(x_1, x_2, x_3)$. For binary "digital" information processing we seek to achieve noise suppression in the vicinity of the logic-point values of the inputs. Assuming approximately equal spread of noise in all the inputs when normalized per their "logic" ranges, the noise-spread transmission factor from the inputs to the output can most in cases be estimated by

$$|\vec{\nabla}z| = \sqrt{\left(\frac{\partial z}{\partial x_1}\right)^2 + \left(\frac{\partial z}{\partial x_2}\right)^2 + \left(\frac{\partial z}{\partial x_3}\right)^2}. \tag{3}$$

The largest value of this quantity when calculated near all the logic points should be less than 1. For each binary gate, the added chemical "filtering" steps can facilitate this.[52,54,55] The function $z(x_1, x_2, x_3; ...)$ depends on physical and chemical parameters (denoted by ...) that are not the scaled inputs $x_{1,2,3}$, but are other quantities that can to some degree be adjusted to modify the system's response. Examples include initial (bio)chemical concentrations of reactants which are not the inputs or output, and various process rates that depend on the chemical and physical conditions.

Not all the optimization tasks can be carried out in the "logic" language, notably, the need to avoid excessive the loss of the overall signal intensity, measured by the spread between $\text{Abs}_{\mathbf{111}}(t = t_g)$ and $\text{Abs}_{\mathbf{000}}(t = t_g)$ values, cf. Equation (2). Signal intensity can be lost due to the added "filtering" processes. Furthermore, the mere possibility of the optimization by "tweaking" the network to change its "analog" information processing responses as a whole or of its constituent "gates," is usually limited to networks that are not too large. For large enough networks "digital" optimization will ultimately be required,[91] involving the redesign of the network with trade-offs involving redundancy, in order to avoid noise buildup.

**NETWORK ELEMENTS: PHENOMENOLOGICAL MODELING**

Modeling of biocatalytic enzymatic processes used as "gates" in multistep processing cascades can be done at various levels. Each enzymatic reaction involves several steps, and can be rather complicated and have various pathways of functioning, some of which are not fully understood and vary depending on the source of the enzyme and other parameters. In our example, for instance (see Figure 1), the mechanism of action of MPh is complicated and not well studied,[80-82,92] whereas GOx has a relatively well understood and straightforward mechanism.[93] HRP has a generally-known, but rather complicated mechanism of action,[94] while HK has a non-unique order of forming complexes with its substrates.[95] In the context of modeling of signal processing networks it is impractical to attempt to fully describe the kinetic processes involved, requiring multiple rate-constant parameters for each of the enzymes.

Not only are the available data not detailed enough for such a description, but it is actually not necessary, as it suffices to describe the function $z(x_1, x_2, x_3)$ semi-quantitatively,[55] in order to evaluate and control its behavior in the vicinity of the logic values of the inputs to improve the network's noise handling[22,31-33,57,58] properties. This can be accomplished by using an approximate, few-parameter fitting for each signal processing step.[54,55] As the network becomes larger, a more engineering approach can be used instead, involving an entirely phenomenological fitting[22,56] that reproduces the general features of $z(x_1, x_2, x_3)$. Ideally a hybrid approach should be favored, with phenomenological fitting expressions derived[56] from simplified kinetic considerations for each sub-process. This offers a connection between the phenomenological fit parameters and controllable physical/chemical properties, enabling adjustments in the network's functioning.[22,56,59]

Let us illustrate these ideas for networked AND gates without filtering. The approach described here[59] also works[56] for an "identity gate" (signal transduction). It uses a Michaelis-Menten (MM) type approximate description[96-98] of enzymatic reactions, with additional approximations for "logic-gate" modeling. A rather surprising result is obtained for parameterizing two-input (two-substrate) AND gates of the type used in our network, Figure 1. We note that AND has generally been the most popular standalone biocatalytic logic gate realized with enzymes.[22,52,55-57,99-102] We use a simplified MM kinetic scheme representing the main pathway for the action of the considered enzyme, $E$,

$$S + E \xrightarrow{k_S} C,  \qquad (4)$$

$$U + C \xrightarrow{k_U} E + P + \cdots,  \qquad (5)$$

where $E$ first binds the substrate, $S$, to form a complex, $C$, that later reacts with the other substrate, $U$, to yield the product, $P$. As common in considering logic-gate design,[55,79] we ignored a possible back-reaction,[103,104] with rate constant $k_{-S}$, in Equation (4), to decrease the number of adjustable parameters, which is possible because in such systems large quantities of substrates are typically used (at least for logic-**1** values) to "drive" the process to yield a large output range. We will further comment on this approximation later.

Enzymatic reactions typically function in an approximate steady state for extended time intervals.[96-98] This need not always be the case; in fact, a very fast reaction regime of saturation was shown to allow avoiding noise amplification in some situations.[58,76] However, the latter regime requires special parameter optimization. In most cases the network parameters are experimentally convenient, but otherwise randomly selected, and we can assume generic behavior for the network's sub-processes. Specifically, in the steady state of a two-input process modeled by Equations (4-5), the fraction of the enzyme that formed the complex is approximately constant; we can assume that

$$\frac{dC}{dt} = k_S SE - k_U UC = k_S SE - k_U U(E_0 - E) \approx 0, \qquad (6)$$

where the subscript 0 denotes values at time $t = 0$. Therefore, in the steady state we expect

$$E \approx \frac{E_0 k_U U}{k_S S + k_U U}, \tag{7}$$

and thus

$$\frac{dP}{dt} = k_U UC \approx \frac{E_0 k_S S k_U U}{k_S S + k_U U}. \tag{8}$$

Since in signal processing applications the reaction is usually driven by plentiful supply of substrates, we can ignore their depletion and write the following approximate expression for the rate of the product production and for its amount at $t = t_g$,

$$\frac{dP}{dt} \approx \frac{E_0 k_S S_0 k_U U_0}{k_S S_0 + k_U U_0}, \tag{9}$$

$$P(t_g) \approx \frac{E_0 k_S S_0 k_U U_0 t_g}{k_S S_0 + k_U U_0}. \tag{10}$$

While several assumptions were made to yield this result,[59] we point out that such expressions are typical of the steady-state-type MM approximations, and were also used to successfully[56,59] fit data. Here we consider a generic two-input AND gate, and therefore the logic-variable response will involve the function $z(x, y)$, with the variables defined according to

$$z = P(t_g)/P(t_g)_{max}, \quad x = S_0/S_{0,max}, \quad y = U_0/U_{0,max}, \tag{11}$$

where the subscript *max* refers to the largest (logic-**1**) values. Substantial parameter cancellations occur as we divide the general Equation (10) by its logic-**1** counterpart, to yield our final expression[59]

$$z(x, y) = \frac{(1+a)xy}{x+ay}, \tag{12}$$

with

$$a = \frac{k_U U_{0,max}}{k_S S_{0,max}}. \tag{13}$$

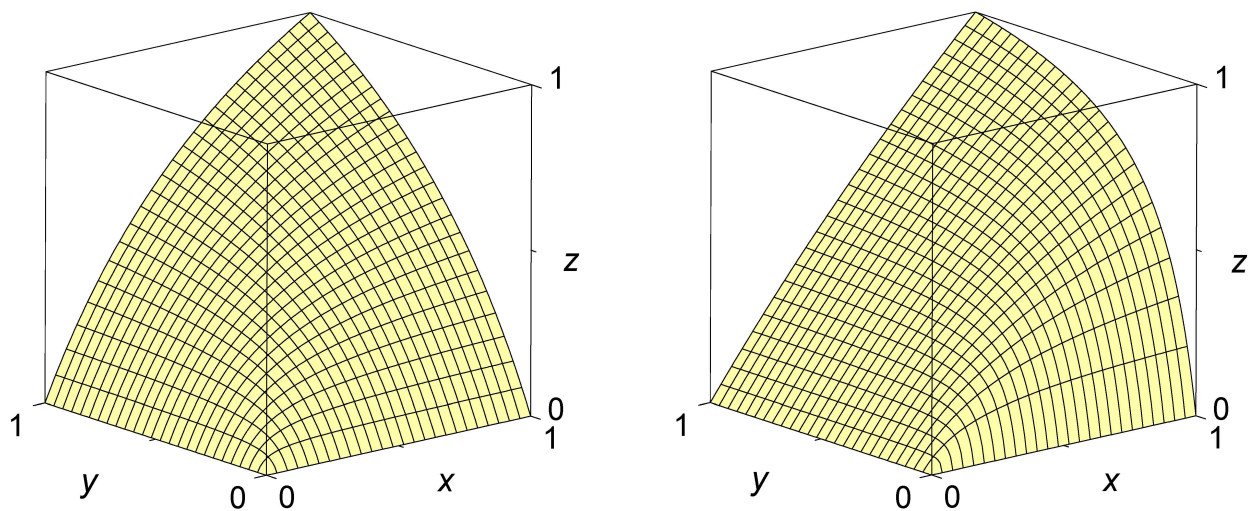

**Figure 2.** The function in Equation (12) for $a = 1$ (left) and $a = 5$ (right).

Interestingly, the logic AND-gate of enzymatic systems in the considered regime can be approximately parameterized with just a *single adjustable parameter*, denoted $a$ in Equations (12-13). This conclusion captures many empirical observations reported earlier for such "non-filtered" gates, when more sophisticated fitting schemes involving kinetic descriptions[54,55] or two-parameter[22] entirely phenomenological expressions were used. It was found[22] that it is difficult to affect the logic-function properties by changing the amount of enzyme or the gate time, which is now explicit in the developed approximations because these quantities ($E_0$ and $t_g$) cancelled out of the expression for $a$ in Equation (13). On the other hand, the logic-**1** values of the two inputs (which are set by the environment in which the gate operates) do affect the shape of the response surface. This is shown in Figure 2, which illustrates a symmetric ($a = 1$) and an asymmetric (here $a > 1$) gate-response functions surfaces described by Equation (12). Note that interchanging the labeling of the inputs, $x \leftrightarrow y$, corresponds to replacing $a \leftrightarrow 1/a$. All such gates are convex and amplify noise, with the noise transmission factor, i.e., the maximal slope of $z(x, y)$ among the four logic points, equal $1 + \max(a, a^{-1})$ in the context of our parameterization. It assumes its smallest value, 2 in the symmetrical case, i.e., 200% noise amplification. This is typical[57] of non-optimized standalone enzymatic AND gates. For asymmetric cases, cf. Figure 2, the noise amplification factor can far exceed 2. Finally, we comment that, with $y = 1$, i.e., with only one varied input, Equation (12) for $z(x)$ reduces to a parameterization[56] of the single-input "identity gate."

We now combine the approximate kinetic expressions of the type shown in Equation (12) for the steps in our example system, Figure 1. In the notation for the logic variables as assigned in the right panel of Figure 1, for each step, except for the last "identity gate" which can be assumed[59] approximately linear here ($z = y_4$), a distinct parameter $a$ is introduced,

$$y_2 = \frac{(1+a_1)x_1 x_2}{x_1 + a_1 x_2}, \tag{14}$$

$$y_3 = \frac{(1+a_2)y_2}{1 + a_2 y_2}, \tag{15}$$

$$z = y_4 = \frac{(1+a_3)x_3 y_3}{x_3 + a_3 y_3}. \tag{16}$$

Concatenating these relations to describe the function $z(x_1, x_2, x_3)$ can be questioned, because the successive steps (gates) feed one another, and therefore intermediate products are time-dependent. However, considering that within the present assumptions the product generation in each step is irreversible, cf. Equation (5), and all the concentrations "driven" by each gate's inputs are linear in the gate-time, Equation (10), the concatenation can be a reasonable approximation:

$$z = \frac{(1+a_1)(1+a_2)(1+a_3)x_1 x_2 x_3}{x_1 x_3 + a_1 a_2 a_3 x_1 x_2 + a_1 a_2 x_1 x_2 x_3 + a_1 a_3 x_1 x_2 + a_2 a_3 x_1 x_2 + a_2 x_1 x_2 x_3 + a_1 x_2 x_3 + a_3 x_1 x_2}. \tag{17}$$

Variants of this expression were used for data fitting.[59] We will further discuss this result after considering the filtering processes.

**INCORPORATION OF BIOCHEMICAL FILTERING IN NETWORKS**

Here we describe modeling[56,59] of added filtering processes. In the literature, there has been recent reports of added "intensity filtering" processes that consume part of an input signal by a competing chemical reaction.[4-6,105] This converts convex biochemical output vs. input response to sigmoid. Other phenomenological descriptions are possible,[5,6,22,32,106-108] notably, the Hill-function fitting[106-108] that is more suitable for sigmoid response caused by cooperativity, when enzyme allostericity or similar effects are considered.[1,2.109,110]

In out example, Figure 1, the first such process is biocatalyzed by enzyme HK, which competes for the input (Glc) of enzyme GOx, and therefore functions as "intensity filtering." The concentration of oxygen is not a varied input, and therefore it can be lumped with the rate constant $k_U$ into a single fixed rate-constant-type parameter combination $\overline{k_U} = k_U U_{0,max}$, cf. Equation (13). The added filtering process biocatalyzed by HK, then competes for a fraction, $F_0$, of the input Glc, up to $S_{0,max}$. Output of that part of the cascade is reduced due to the diversion of part of the input, and this can be phenomenologically modeled in a simplified fashion by adding the process

$$S + F \xrightarrow{k_F} \ldots, \tag{18}$$

where $F$ is initially set to $F_0$. The parameters $F_0$ and $k_F$ are phenomenological, because this is only an approximate description of the added HK step (Figure 1). For the considered case, $F_0$ can be approximately adjusted by varying the initial concentration of ATP, whereas the overall process rate constant, lumped in $k_F$, can be varied by changing the amount of HK. This approximation aims at obtaining a simple fitting expression: Equation (18) alone, when written as a rate equation for $S(t)$, suggests depletion of the availability of the substrate $S$ at the gate time $t_g$ according to

$$S(t_g) = (S_0 - F_0)S_0/[S_0 - F_0 e^{-k_F(S_0-F_0)t_g}] \,. \tag{19}$$

We then use[56,59] this expression as accounting for the reduced intensity, to replace $S_0$ in Equation (10), with $U_0$ set to $U_{0,max}$, to write

$$P(t_g) \approx \frac{k_S k_U U_{0,max} E_0 t_g}{k_S + \frac{k_U U_{0,max}}{(S_0-F_0)S_0}[S_0 - F_0 e^{-k_F(S_0-F_0)t_g}]} \,. \tag{20}$$

In terms of the scaled variables for this step, see Figure 1, and its earlier introduced parameter $a = a_2$, we then obtain[59] the following expression to replace Equation (15),

$$y_3(y_2) = \frac{y_2(y_2-f_2)\{1-f_2+a_2[1-f_2 e^{-b_2(1-f_2)}]\}}{(1-f_2)\{y_2(y_2-f_2)+a_2[y_2-f_2 e^{-b_2(y_2-f_2)}]\}} \,. \tag{21}$$

Except for relabeling the scaled variables and adding index 2 to the fitting constants to designate the gate, this is essentially the same expression as derived in earlier work,[56] with the general relations for the new fitting parameters (without the index 2),

$$f \equiv F_0/S_{0,max}, \quad b \equiv k_F S_{0,max} t_g \,. \tag{22}$$

Note that we expect the values of the parameters defined in this section to satisfy

$$a > 0, \quad 0 \leq f < 1, \quad b \geq 0. \tag{23}$$

For individual gates, added filtering processes frequently improve noise-transmission properties by making their response sigmoid in one or both inputs. The parameter $f$ can be adjusted by varying the amount of the supplied "filtering" chemical (here, ATP), whereas the parameter $b$ can be changed not only by varying the process rate (here, by amount of HK) but also by selecting the gate time, $t_g$, cf. Equation (22). Plots of functions[56] such as Equation (21) for representative parameter values were illustrated in Figure 3.

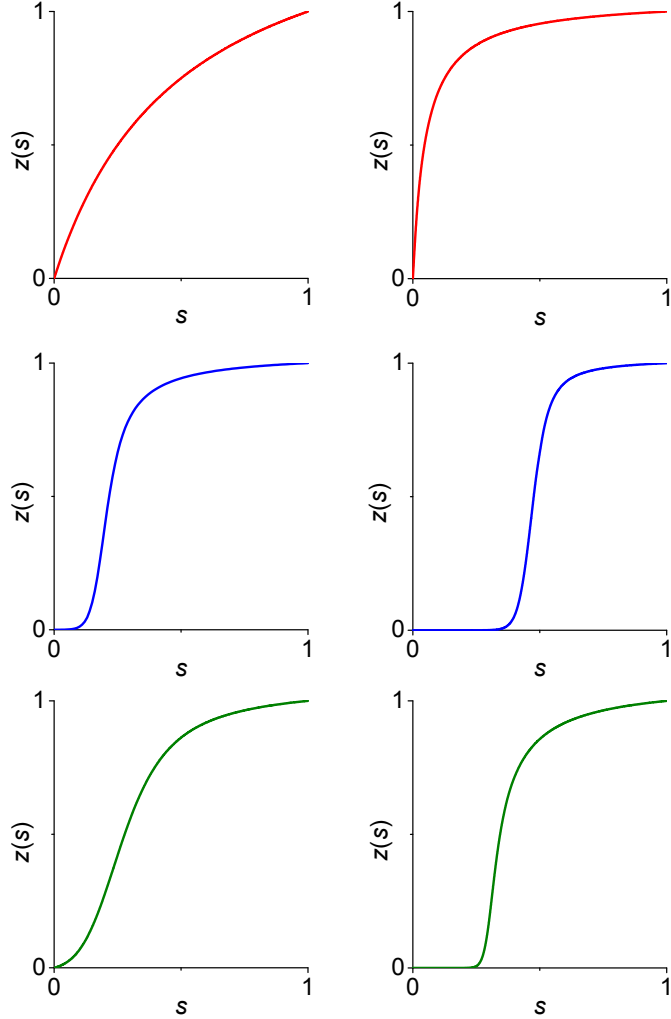

**Figure 3.** Sketches of the fitting functions with parameter values randomly selected for illustrations. The top panels show the convex function such as the one in Equation (15), in terms of a generic notation $s$ for the input signal, and $z(s)$ for the output: $z(s) = (1 + a)s/(1 + as)$, illustrating, from left to right, the effect of decreasing the value of $a$. The middle panels plot a sigmoid function, such as in Equation (21), $z(s) = s(s - f)\{1 - f + a[1 - fe^{-b(1-f)}]\}/\big((1 - f)\{s(s - f) + a[s - fe^{-b(s-f)}]\}\big)$, with, from left to right, increasing the parameter $f$, with the other two parameters, $a$ and $b$, fixed. The bottom panels show the effect of increasing the parameter $b$, with the other two parameters fixed (and not related to the values use in the middle panels).

We now consider an added output-filtering process for two-input AND gates. In the present system, the third (HRP) step of the processing, see Figure 1, has one such an added chemical filter that chemically "recycles" the output into one of the input substrates, TMB, by the added NADH, as long as the latter is not used up. We bypass the difficulty of modeling it directly, by considering it as a part of the network in which, as shown in Figure 1, we in advance somewhat artificially singled out the chemical-to-optical signal conversion as an additional *single*-input "identity gate." We consider the added filter process as competing for the input, $TMB_{ox}$, of this step, which was earlier regarded as approximately linear. We note that linear response is obtained as the limit of large $a$ in our phenomenological modeling of single-input identify functions, cf. Equation (15) for a different step. Therefore, we adopt the $a \to \infty$ limiting form of the expressions with filtering, such as Equation (21), instead of the final-step linear function, see Equation (16), i.e., we take

$$z(y_4) = \frac{y_4(y_4-f_3)\left[1-f_3 e^{-b_3(1-f_3)}\right]}{(1-f_3)\left[y_4-f_3 e^{-b_3(y_4-f_3)}\right]}, \tag{24}$$

but the relation for $y_4(y_3)$ in Equation (16) remains unchanged. Here subscript 3 designates the two added fitting parameters, $f_3$ and $b_3$, of to the filtering process involving NADH reacting with the output of the third gate in the original cascade, consistent with the notation for $a_3$ for that gate. The parameter $f_3$ can be approximately adjusted by varying the NADH concentration, whereas $b_3$, related to the rate constant, can be changed by adjusting the gate time.

Various relations derived in this section can be concatenated to write down expressions which replace the "no filters" Equation (17) with appropriate formulas for the cases of one or both of the filtering processes shown in Figure 1 added. This is summarized in Figure 4. These analytical expressions are too cumbersome to display explicitly. However, we point out that the concatenation can be done in a computer, and the whole network description is easily programmed for data fitting.[59]

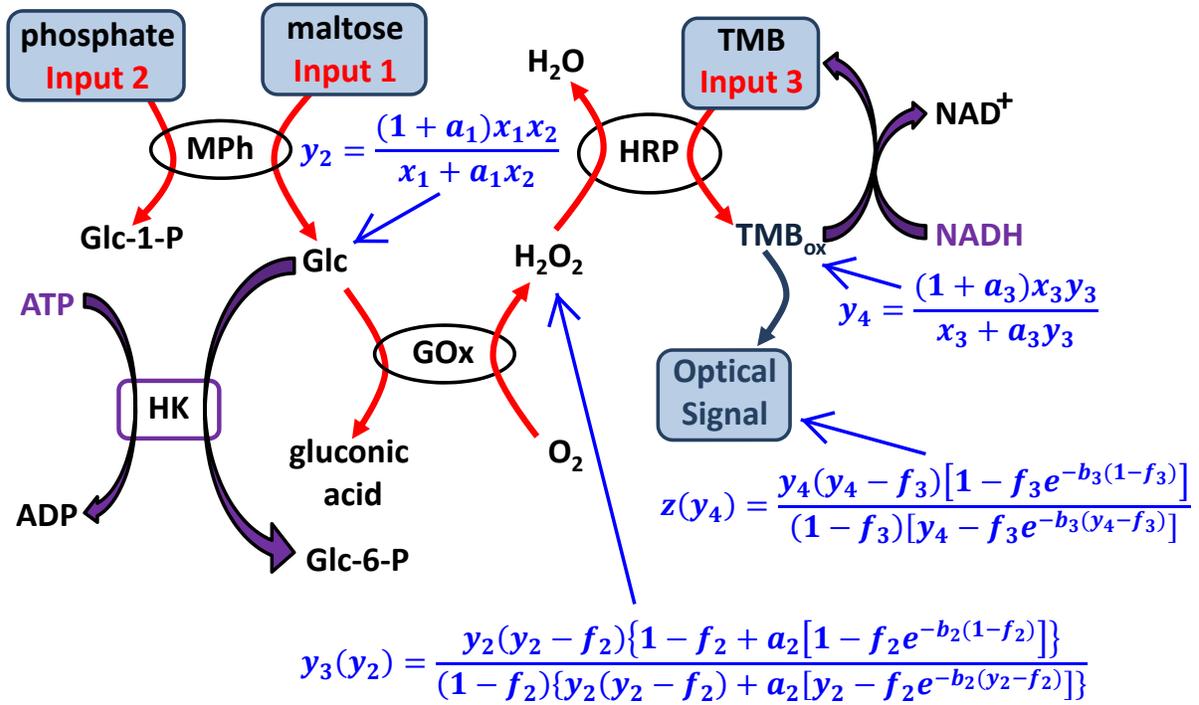

**Figure 4.** Enzymatic cascade[59] introduced in Figure 1, with various relations derived in the text shown for the appropriate intermediate and final signals, see Equations (14), (16), (21), (24).

**CONCLUSION**

In this work we reviewed a new single-parameter parameterization for two-input enzymatic AND gates without filtering, Equation (12), which captures several known properties of such processes. We then described a recently developed approach[56,59] to adding the filtering processes by phenomenologically modeling the resulting systems using closed-form analytical expressions, e.g., Equations (21), (24). The phenomenological functions worked reasonably well in fitting experimental data sets[56,59] to determine the parameters in groups of one or two at a time, as well as later reproducing other measured data sets with the fitted parameters, without any additional adjustments. It transpires that, scaled (to reference ranges) "logic variables" for the inputs, output and some intermediate products can be useful in describing enzyme cascades by identifying quantities that offer the most direct control of the network properties, and also allowing to approximately fit the system's responses with few adjustable parameters.

ACKNOWLEDGEMENTS

The author thanks his numerous colleagues,[1,2,3,11,22,23,31-37,47,52-59,76-79,91,99] notably, Prof. E. Katz, for collaborative team work. He gratefully acknowledges funding of the work reviewed here by the US National Science Foundation, most recently under grant CBET-1403208.